\documentclass[twocolumn]{aastex6}
\usepackage{gensymb}
\usepackage{natbib}
\begin{document}
\turnoffedit{}
\title{ALMA Thermal Observations of Europa}

\author{Samantha K. Trumbo and Michael E. Brown}
\affil{Division of Geological and Planetary Sciences, California Institute of Technology, Pasadena, CA 91125, USA}
\author{Bryan J. Butler}
\affil{National Radio Astronomy Observatory, Socorro, NM 87801, USA}

\begin{abstract}
We present four daytime thermal images of Europa taken with the Atacama Large Millimeter Array. Together, these images comprise the first spatially resolved thermal dataset with complete coverage of Europa's surface. The resulting brightness temperatures correspond to a frequency of 233 GHz (1.3 mm) and a typical linear resolution of roughly 200 km. At this resolution, the images capture spatially localized thermal variations on the scale of geologic and compositional units. We use a global thermal model of Europa to simulate the ALMA observations in order to investigate the thermal structure visible in the data. Comparisons between the data and model images suggest that the large-scale daytime thermal structure on Europa largely results from bolometric albedo variations across the surface. Using bolometric albedos extrapolated from \textit{Voyager} measurements, a homogenous model reproduces these patterns well, but localized discrepancies exist. These discrepancies can be largely explained by spatial inhomogeneity of the surface thermal properties. Thus, we use the four ALMA images to create maps of the surface thermal inertia \edit1{and emissivity at our ALMA wavelength}. From these maps, we identify a region of \edit1{either} particularly high thermal inertia \edit1{or low emissivity} near 90 degrees West and 23 degrees North, which appears anomalously cold in two of our images. 
\end{abstract}

\keywords{planets and satellites: general --- planets and satellites: individual (Europa) --- planets and satellites: surfaces}

\section{Introduction}\label{sec:intro}
Europa's icy surface is marked by fractured, ridged, and chaotic terrain suggestive of a history of geologic activity \citep[e.g.][]{Kattenhorn2009, Prockter2009, Collins2009}. Spectroscopic studies have revealed multiple compositions that reflect the influences of both endogenous geologic processes \citep[e.g.][]{McCord1998, Fischer2015} and exogenous radiolytic processing by Jovian magnetospheric particles \citep[e.g.][]{Carlson1999, Carlson2002}, but the balance of these influences in shaping surface properties is not well understood. Surface temperature measurements can provide an additional window onto both types of processes. Such measurements present perhaps the best means for identifying regions of active geologic activity. Indeed, active hotspots persist at both the ``tiger stripes" of Enceladus \citep{Spencer2006} and the volcanoes of Io \citep{Pearl1982,Spencer1990Io}. In addition, \textit{Cassini} thermal observations of Saturn's moons Mimas and Tethys have shown that temperature measurements can reveal details on the effects of magnetospheric particle bombardment on surface texture \citep{Howett2011Mimas, Howett2012Tethys}. Finally, thermal data can provide insight on diurnal sublimation cycles, impact gardening by micrometeorites, and sputtering from particle impacts, which also affect the surface compositions and morphologies and, thus, the surface thermophysical properties.

To date, the only spatially resolved thermal data of Europa were collected by the \textit{Galileo} Photopolarimeter-Radiometer (PPR). These data provided the first brightness temperature maps of the surface and included both daytime and nighttime measurements \citep{Spencer1999}. Modeling efforts using the PPR data have found thermal inertia values \edit1{between 30 -- 140 $J/(m^2 \cdot K \cdot s^{1/2})$,} consistent with a particulate, uncompacted regolith texture unlike that of solid water-ice \citep{Spencer1999,Rathbun2010}. However, the \textit{Galileo} PPR only obtained limited coverage of the surface. Furthermore, from the end of the \textit{Galileo} mission until very recently, subsequent brightness temperature measurements of similar quality and spatial resolution have been impossible to achieve. Recently, however, the Atacama Large Millimeter Array (ALMA) has made the collection of spatially resolved, high-quality thermal datasets possible. 

Here, we present four ALMA thermal images that together cover the entire surface of Europa at a frequency of 233 GHz (1.3 mm) with a typical linear resolution of $\sim$200 km. Using a global thermal model of Europa, we fit the observations and investigate the nature of thermal structure visible across the surface.

\section{ALMA Observations and Data Reduction}\label{sec:methods}
The observations of Europa were taken with the main array of the Atacama Large Millimeter Array (ALMA), which is composed of up to 50 12-meter diameter antennas spread across the Altiplano in the high northern Chilean Andes.  Every pair of antennas acts as a two-element interferometer, and together these individual interferometers allow for the reconstruction of the the full sky brightness in both dimensions
\citep{2001_Thompson}.

ALMA can operate in 7 frequency windows, from $\sim$90 to $\sim$950 GHz. The observations presented here were taken in Band 6, near 230 GHz, in the ``continuum'' (or ``TDM'') mode, with the standard frequency tuning.  This results in four spectral windows with frequencies: 223--225 GHz; 225--227 GHz; 239--241 GHz; and 241--243 GHz. In the data reduction, we average over the entire frequency range, and use 233 GHz as the observation frequency in our thermal modeling.

We observed Europa with ALMA on November 17, 25, 26, and 27 of 2015. For these observations, there were between 38 and 43 antennas, in the C36-7 configuration.  This configuration has a maximum antenna spacing of $\sim$5 km, giving a resolution on the sky of $\sim$50 milliarcseconds\edit1{, and a minimum antenna spacing of $\sim$250 m, giving a largest theoretical recoverable scale of $\sim$0.65 arcseconds}.  Details of observation times, geometries, and resolutions are given in Table 1.

\floattable
\begin{deluxetable}{cccccc}
\tablecaption{Table of Observations\label{table:observations}}
\tablecolumns{3}
\tablenum{1}
\tablewidth{0pt}
\tablehead{
\colhead{Date} & \colhead{Time}  & \colhead{Sub-Earth Longitude} & \colhead{Sub-Earth Latitude} &\colhead{Elliptical Beam Resolutions} \\
\colhead{(UT)} & \colhead{(Start/End)} & \colhead{Range} & \colhead{} & \colhead{(Milliarcseconds)}}

\startdata
2015 Nov 17 & 14:36/14:50 &  44.9--45.9 & -1.44 & 57/42 \\
2015 Nov 25 & 11:25/11:41 &  121.5--122.6 & -1.52 & 59/53\\
2015 Nov 26 & 08:22/08:38 &  210--211.1 & -1.52 & 77/52\\
2015 Nov 27 & 10:00/10:40 &  318.1--320.9 & -1.54 &  54/53\\
\enddata
\end{deluxetable}

All of these observations are in dual-linear polarization; in the end we combine these into a measurement of Stokes I.  While we expect  polarized emission from the surface, it is relatively weak, and we did not measure the cross-polarized signals.

% number of antennas: 43, 42, 38, 38

Each observation was about 20 minutes in duration, including all calibration overheads, which resulted in roughly 8 minutes on Europa. The point-like calibrator J1058+0133 was used as the absolute flux density scale calibrator for all four observations --- it is part of a grid of calibrators which are regularly monitored against the main flux density scale calibrators for ALMA \citep{2012_Butler}.  The nearby calibrator J1108+0811 was used to calibrate the phase of the atmosphere and antennas as a function of time.

Initial calibration of the data was provided by the ALMA observatory and completed in the CASA reduction package \citep{McMullinEtAl2007} via the ALMA pipeline \citep{2014_Muders}.  We exported the provided visibilities from CASA and continued the data reduction in the AIPS reduction package \citep{aips}.  We
self-calibrated \citep{1999_Cornwell} the data in three iterations down to a time interval of 8 seconds (possible because Europa is such a bright target), using an initial limb-darkened model and imaging more deeply in each iteration \citep{1999_Butler}. \edit1{The initial model in each iteration is parameterized to account for limb darkening and includes the diameter (taken to be known from the observing geometry), the total flux density, and and the limb-darkening parameter, where both the flux density and the limb-darkening parameter are taken from actual fits to the visibilities at each step using the AIPS task OMFIT.} 

\edit1{Inspection of the resulting images from the four days of observation indicated that there was an apparent offset in the overall brightness of the November 17 data compared to the other days --- the November 17 data was less bright by $\sim$10\%. Examination of the derived flux densities for J1108+0811 confirmed this offset. To determine the cause of the offset, we searched the ALMA grid calibrator monitoring catalog for flux densities derived for J1058+0133 (our flux density scale calibrator) over the period of time relevant to our observations and compared them to the flux densities used in the ALMA data reduction pipeline. We found that, while the cataloged flux densities at our frequency show the source decreasing in brightness from 3.42 $\pm$ 0.07 Jy on November 17 to 3.02 $\pm$ 0.10 Jy on November 25, this decrease was not properly reflected in the values used in the data reduction pipeline (3.37 Jy for November 17 and 3.33 Jy for November 25-27). We therefore made corrections to the flux density scale of the visibilities (and hence the brightness temperatures in the images) by accounting for the flux density differences between the measurements of J1058+0133 from November 17 and November 25 and the values used in the pipeline data reduction. The resultant correction factors are 1.015 for November 17 and 0.907 for the remaining three dates.}

The final images are shown in Figure 1. They were produced using a robust parameter of 0, which is a good compromise between resolution and signal-to-noise. Pixel sizes are roughly a factor of 10 smaller than the actual resolution to minimize deconvolution errors. At the time of our observations, Europa's projected diameter was $\sim$0.77$^\prime$$^\prime$ on the sky, resulting in $\sim$15 resolution elements across the disk. \edit1{We note that this diameter is larger than the theoretical largest recoverable scale noted above. However, as the structure of the visibilities (and the sky brightness itself) is well-known for circular sources, and, as we use fits of the visibilities and an initial model incorporating the fitted zero-spacing flux density as the first step in all of our imaging, the overall brightness level is well-constrained \citep{1999_Butler}.}

\section{Thermal Modeling}\label{sec:modeling} 
We use a global thermal model of Europa, similar to those used for other Solar System bodies \citep[e.g.][]{Spencer1989, Spencer1990, Hayne2015}, to simulate the ALMA observations. The model begins by calculating the absorbed solar flux at each point on the surface according to the local bolometric albedo. The resulting heat is then conducted through the near-surface layers, where the temperatures as a function of depth and time are controlled by the thermal inertia. The model numerically solves for these temperatures using a small global heat flux of 20 $mW/m^2$ \citep{MitriShowman2005, BarrShowman2009} as a lower boundary condition and grey body radiation to space as an upper boundary condition. The end result is a radiative flux map of the surface that can be output for the geometries and times of the ALMA observations, smoothed to the ALMA resolution, and converted to brightness temperature. A complete description of the model can be found in \citet{Trumbo2017alma}.

The primary difference in our modeling approach from those taken in the past for Europa \citep[e.g.][]{Spencer1999, Rathbun2010} is that we do not fit for bolometric albedo, but rather use an albedo map constructed from discrete \textit{Voyager} measurements. We take the normal albedos for distinct locations on the surface from \citet{McEwen1986} across the green, blue, violet, and ultraviolet Voyager filters. We then weight the values in each filter by the solar flux in the corresponding wavelength range and average the values. \edit1{As the phase integral of Europa is 1.01 \citep{Grundy2007}, we take these averages as approximate bolometric albedos at each location.} Finally, we take these as tie-points to the greyscale \textit{Voyager}/\textit{Galileo} global mosaic of Europa \citep{USGSmap}. We fit a line to the relationship between the pixel values in the image at the Voyager albedo locations and the approximate bolometric albedo tie-points, and use this relationship to produce a high-resolution albedo map from the mosaic. The pixel values and albedo tie-points correlate with an R$^2$ of 0.92 and a standard deviation in albedo of 0.03. We use the resulting map in our thermal modeling and treat the surface thermal inertia and emissivity \edit1{at our ALMA wavelength} as free parameters, \edit1{assuming a fixed bolometric emissivity of 0.9 \citep{Spencer1987}.}

\edit1{Bolometric} emissivities less than one will increase the physical temperatures of the surface by inhibiting radiative heat loss. However, as ALMA observes brightness temperatures at 233 GHz (1.3 mm), rather than physical temperatures, a decrease in emissivity \edit1{at ALMA wavelengths} actually tends to decrease the observed brightness temperatures everywhere on the disk for the same reason. Variations in thermal inertia, on the other hand, affect temperatures differently depending on the time of day. An increase in thermal inertia will flatten the diurnal temperature curve by reducing the contrast between nighttime and midday temperatures. In contrast, a decrease in thermal inertia will steepen the diurnal temperature curve, lowering nighttime temperatures and increasing them at midday. 

When fitting the data using our simple thermal model, however, the deduced thermal inertia and emissivity may also include minor contributions from physical processes not included in the model. In particular, our model does not include sunlight penetration with depth in the regolith, thermal emission from depth in the regolith, or surface roughness. As described in \citet{Trumbo2017alma}, our model assumes that the solar flux is absorbed in the topmost model layer, which is a standard assumption for several thermal models \citep[e.g.][]{Spencer1989, Spencer1990, Hayne2015} and particularly valid for low-albedo surfaces. However, sunlight may penetrate deeper into a relatively high-albedo particulate surface like that of Europa \citep{Brown1987, Urquhart1996}, resulting in heat at depth in the regolith. In practice, this effect is difficult to distinguish from that of thermal inertia \citep{Urquhart1996}, thus we include only thermal inertia in our model. 

Our model also assumes that the thermal emission detected by ALMA originates in the topmost model layer. At a frequency of 233 GHz (1.3 mm), however, ALMA is likely sensing slightly deeper into the surface. Thus, modeled brightness temperatures for a fixed emissivity and thermal inertia are slightly higher than they would be if this effect were included. We find that the inclusion of sunlight absorption at depth (at an e-folding of 2 cm) and thermal emission from depth (with an e-folding of 1 cm) results in slight changes to the model thermal inertia and emissivity, respectively, but has little effect on our conclusions from the data. Similarly, the inclusion of surface roughness appears to have little effect on our results. Surface roughness tends to increase surface temperatures, particularly at the limbs. However, using a similar implementation of surface roughness to \citet{Hayne2015}, we find that rms slopes up to 20$\degree$ have relatively minor effects on our model fits to the data. 

\deleted{Finally, our model assumes an emissivity that is constant with wavelength, as the relevant emissivities of Europa-like compositions under Europa-like conditions are unknown. Thus, our model emissivity acts both as the bolometric emissivity, which modulates the efficiency of grey body radiation to space, and as the emissivity controlling brightness temperatures at the ALMA wavelength. If we were to instead separate the two parameters, then an increase in bolometric emissivity would decrease physical surface temperatures in the model by increasing the efficiency of radiative heat loss from the top layer. In order to maintain the same brightness temperatures, this change would need to be balanced by an increase in the emissivity at our ALMA wavelength. The two decoupled emissivities may both be different than the value deduced when the two parameters are combined, but the effect on our fits to the data would be small and approximately independent of geographic location.}

In our analysis, we aim to explain the nature of the large-scale thermal structure of Europa and search for any potential systematic variation of thermal properties across the surface, rather than precisely determine the true values of the individual surface thermal properties. Thus, we use our simplest model and note that our model parameters may reflect the influences of the above effects.

\deleted{We would like to point out one final complicating factor in modeling the data and interpreting the thermal structure. Due to inherent variability in the flux output of calibrator sources used by ALMA (mostly quasars), there may be time-dependent uncertainties in the absolute flux calibration of ALMA images, and thus in their brightness temperatures. Therefore, relative flux discrepancies may exist between images taken on different dates. This effect introduces an additional potential parameter for each image, which can complicate the inference of thermal properties from the data. Without repeat spatial coverage at the same time of day, such flux errors can be difficult to separate from spatial variability in surface thermal properties. In our case, it is only after significant preliminary analysis that the likelihood of flux discrepancies becomes apparent. We walk through all of the steps in our preliminary and final analyses in Section \ref{sec:fits}, but would like to emphasize that flux adjustments become important and that our figures reflect our final analysis, including these adjustments.}

\section{Fits to ALMA Observations}\label{sec:fits} 
We begin our analysis by attempting to determine a global best-fit emissivity and thermal inertia. We simulate the four ALMA observations over a grid of thermal inertias and emissivities and find the single best-fit values to these data by minimizing the square of the residuals between model and data images. This initial fitting produces a global best-fit emissivity of \replaced{0.77}{0.75} and a best-fit thermal inertia of \replaced{85}{95} $J/(m^2 \cdot K \cdot s^{1/2})$. This thermal inertia is higher than the value of 70 $J/(m^2 \cdot K \cdot s^{1/2})$ found by \citet{Spencer1999} for the equatorial regions, but lies within the range of 30 -- 140 $J/(m^2 \cdot K \cdot s^{1/2})$ mapped by \citet{Rathbun2010}.

\begin{figure*}
\figurenum{1}
\plotone{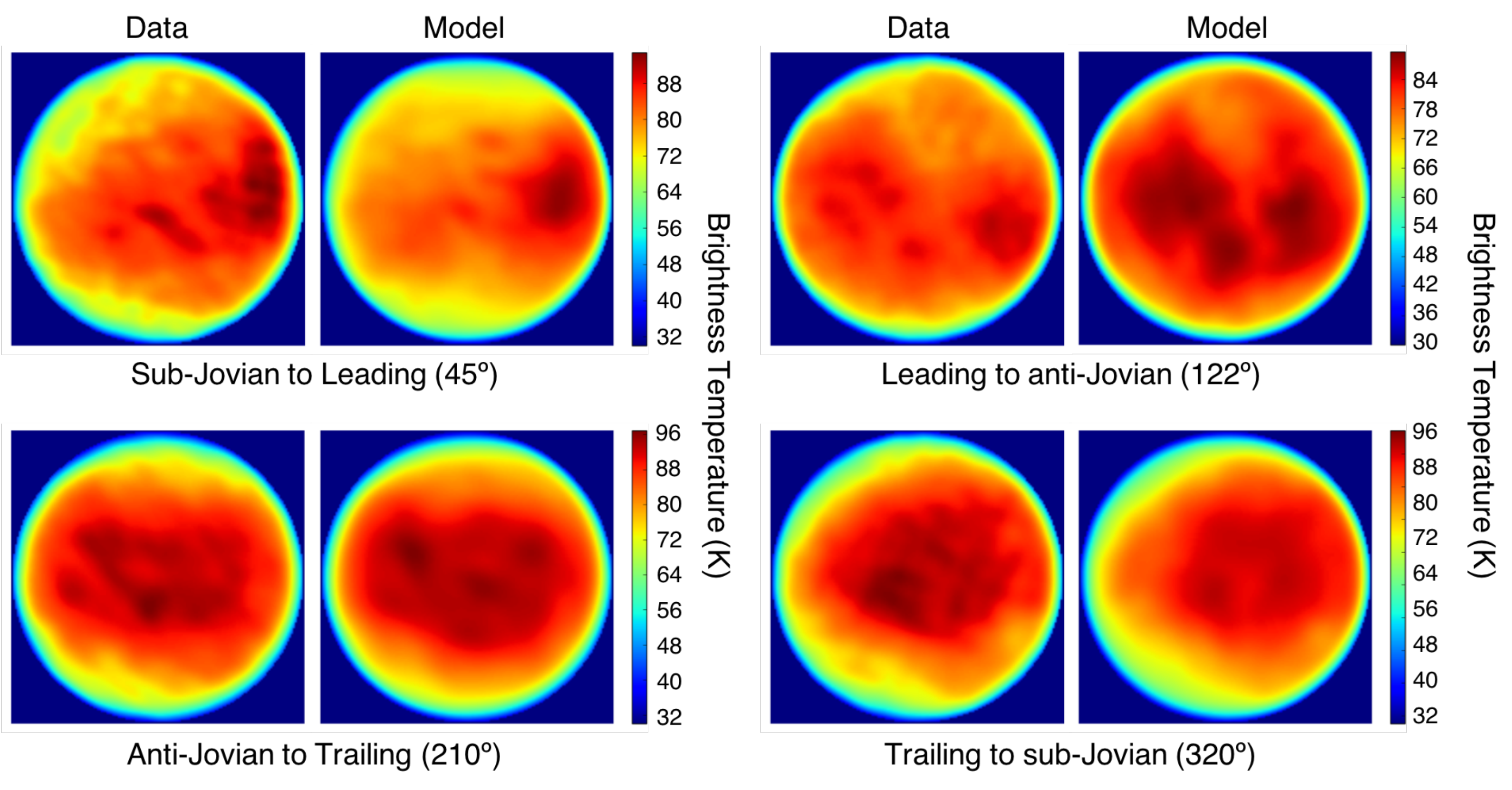}
\caption{Comparison of ALMA images with model images produced using a globally homogenous thermal model and our best-fit emissivity of 0.75 and thermal inertia of 95 $J/(m^2 \cdot K \cdot s^{1/2})$. Large-scale thermal structure is well-reproduced and many corresponding features are identifiable in each data-model pair. \label{fig:datamodel}}
\end{figure*}

\begin{figure}
\figurenum{2}
\plotone{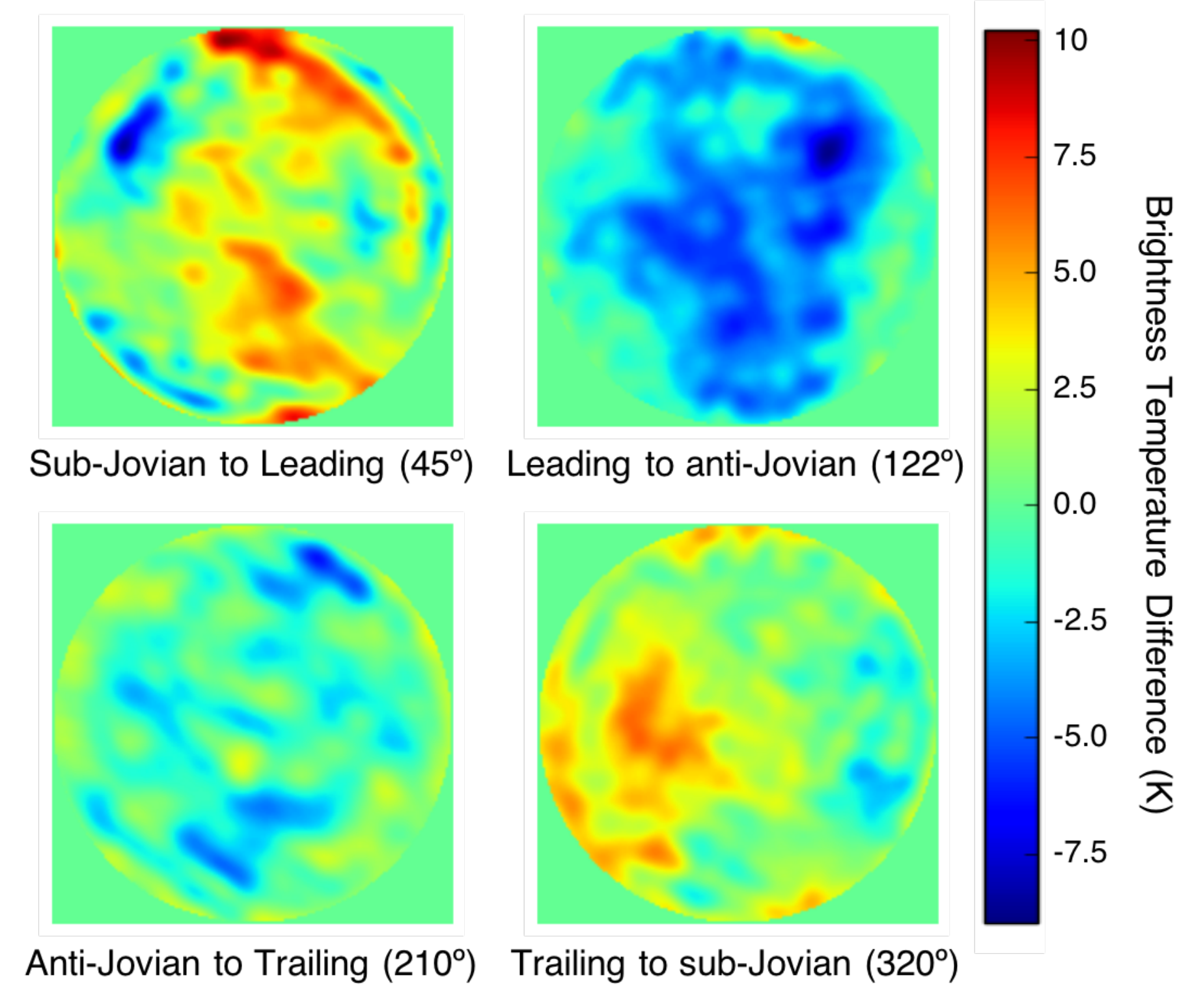}
\caption{Residuals between the data and model images produced with our best-fit parameters, where positive temperatures indicate places where the data are hotter than the model predicts. Discrepancies may result from local variations in the surface thermal properties. \label{fig:residuals}}
\end{figure}

Data images are shown in Figure \ref{fig:datamodel} alongside model images produced using these global best-fit parameters. The globally homogenous model, where only albedo varies spatially across the surface, reproduces the large-scale thermal structure of the data well. This suggests that the majority of the visible daytime thermal structure on Europa is governed primarily by local albedo variations, rather than variations in internal heating, thermal inertia, or emissivity. However, when we difference the model and data images, localized discrepancies between the data and model do become apparent. Figure \ref{fig:residuals} shows the residuals between each data-model image pair from Figure \ref{fig:datamodel}, where positive temperatures indicate regions where the data are hotter than the model predicts. Such discrepancies are not necessarily surprising given the inhomogeneous nature of Europa's surface, and we do not expect the entire surface, with its varying compositions and morphologies, to be well-represented by a single thermal inertia or emissivity. 

In principle, discrepancies may be due to a number of possible causes. Localized differences in the thermal inertia, emissivity, or albedo from the values used in our model will result in residuals on the spatial scale of the geographic variability in these properties. Other possible contributing factors include spatial variation in the transparency of the surface to sunlight or thermal emission, which may manifest as apparent thermal inertia or emissivity discrepancies in our modeling. As we lack observations \edit1{of the same regions} at multiple times of day over most of the surface, we cannot conclusively distinguish between these potential causes everywhere. \edit1{For instance, most of the positive temperature anomalies in the sub-Jovian to leading hemisphere image of Figure \ref{fig:residuals} can be equivalently explained by a decrease in albedo of $\sim$8--20\% or an elevated subsurface heat flux of $\sim$1.2 -- 2 $W/m^2$ (consistent with liquid water a few hundred meters below the surface).} However, as we provide the model with spatially varying albedos derived from concrete measurements and because other objects in the Solar System exhibit significant localized differences in thermal inertia \citep[e.g.][]{Howett2010, Howett2011Mimas, Howett2012Tethys, Hayne2017, Putzig2005}, we experiment with the idea that the residuals may be caused by thermal inertia variations. 

\begin{figure*}
\figurenum{3}
\plotone{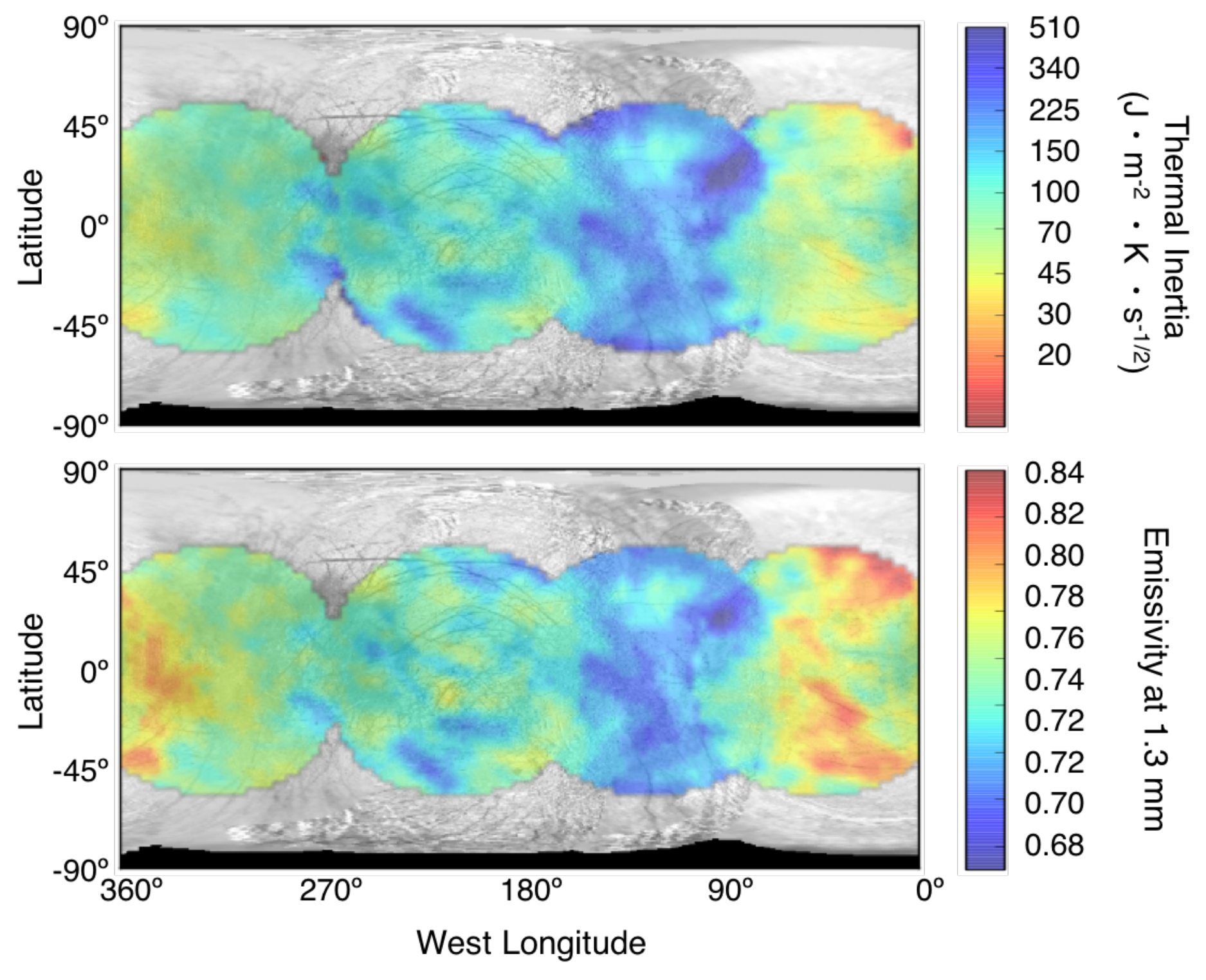}
\caption{Top: Thermal inertia map created by assuming a fixed emissivity of 0.75 (our global best-fit value) and allowing thermal inertia to vary at each point across the surface. Note the elevated thermal inertia near Pwyll crater \edit1{(271$\degree$ W and 25$\degree$ S)} and the anomalously high thermal inertia at 90$\degree$ W and 23$\degree$ N, the location of the recurring cold spot in our ALMA data. \edit1{Bottom: Millimeter emissivity map created assuming a fixed thermal inertia of 95 $J/(m^2 \cdot K \cdot s^{1/2})$ (our global best-fit value) and allowing emissivity to vary spatially across the surface.} The background basemap is from USGS Map-a-Planet: https://www.mapaplanet.org/explorer/europa.html. \label{fig:map}}
\end{figure*}

We fix the emissivity at the global best-fit value of 0.75 and fit the data by letting the thermal inertia vary at each point. Points near the limbs of each observation are \edit1{foreshortened, convolved with the cold sky, and more sensitive to surface roughness, positioning uncertainty, and other effects that are unimportant away from the limb. In addition, these locations are experiencing times of day near where diurnal curves for varying thermal inertias (at a given emissivity) intersect. The combination of this effect with the magnified uncertainties near the limbs can result in widely varying best-fit thermal inertias that are discontinuous with those of surrounding areas.} Thus, we only fit data within 57$\degree$ of the central point of each observation, which we find minimizes this effect without compromising our longitudinal coverage. In fitting regions that appear in two overlapping images, we use both times of day in the fitting. \edit1{This produces a map of thermal inertia of the surface (Figure \ref{fig:map}), assuming a globally homogenous emissivity and our bolometric albedo map constructed from \textit{Voyager} measurements.} \deleted{At this point, flux calibration offsets between observations become apparent. When we take this approach, without flux correcting the images, we see obvious inconsistencies in overlapping portions of our data. In particular, regions of overlap between the 17 Nov. observation (sub-Jovian to leading hemisphere) and those on either side appear anomalously cold in the 17 Nov. image, resulting in poor fits at both times of day and best-fit local thermal inertias that are drastically discontinuous with surrounding areas. The 17 Nov. observation was separated from the other three by at least a week and has anomalously low fluxes relative to the remaining three images, which were taken on consecutive days. This, along with the substantial discrepancies in regions of overlap, suggests a nontrivial flux calibration offset for this image, implying that our preceding analysis must be revisited while accounting for this possibility.}

\deleted{In order to put the data on a common flux scale, we fit for flux calibration adjustments using overlapping portions of our data. We utilize the methodology outlined above, where we fix the emissivity and let thermal inertia vary at each point. However, we can no longer take an emissivity of 0.77 as the global best fit, because it was derived using the anomalously cold Nov. 17 observation. Instead, we redo the global best-fit using using only the three consistent observations and arrive at a similar best-fit emissivity of 0.80 and thermal inertia of 100 $J/(m^2 \cdot K \cdot s^{1/2})$. Then, fixing the 26 Nov. observation with no flux adjustment, we sequentially fit for flux calibration discrepancies of the remaining images using the overlapping regions of the 26 and 25 Nov. images, the 25 and 17 Nov. images, and the 17 and 27 Nov. images, while letting thermal inertia vary independently at each point on the surface. We find a best-fit flux calibration offset of 12\% for 17 Nov. (sub-Jovian to leading hemisphere), 2\% for 25 Nov. (leading hemisphere to anti-Jovian), and -1\% for 27 Nov. (trailing hemisphere to sub-Jovian). We then redo the global best fit using all four images with the flux adjustments applied, converging on the same best-fit emissivity of 0.80 and a similar thermal inertia of 90 $J/(m^2 \cdot K \cdot s^{1/2})$. We take these values as our final global best-fit parameters. Again, this thermal inertia is higher than that found by Spencer et al. 1999 using \textit{Galileo} data, but within the range of values Rathbun et al. 2010 mapped using the same data.}

\deleted{The data images of Figure 1 include the above flux corrections, and the adjacent model images are produced using these final best-fit parameters. Our major conclusion remains unchanged---the large-scale thermal structure visible in the Europa images is primarily controlled by bolometric albedo variations across the surface, rather than differences in local heat flow, thermal inertia, or emissivity. Still, however, localized discrepancies between the data and the homogenous model exist (Figure 2). As discussed above, such discrepancies may be due to a number of potential causes, which are indistinguishable without measurements at multiple times of day. For instance, most of the positive temperature anomalies in the sub-Jovian to leading hemisphere image of Figure 2 can be equivalently explained by a decrease in albedo of $\sim$ 8--20\% or an elevated subsurface heat flux of $\sim$ 1.2--2 $W/m^2$ (consistent with liquid water a few hundred meters below the surface). However, thermal inertia inhomogeneities can also explain most of these anomalies.}

\deleted{As a final step, we again experiment with the idea that these corrected residuals are due to local variations in the surface thermal inertia. We now fix the emissivity at the final best-fit value of 0.80 and again fit the data by letting the thermal inertia vary independently at each point, where both times of day are used in fitting regions of overlap. This produces a map of the thermal inertia of the surface (Figure 3), assuming a globally homogenous emissivity and our bolometric albedo map constructed from \textit{Voyager} measurements.}

Under the assumption that all of the residuals between the ALMA data and the global best-fit model can be attributed to thermal inertia variations, typical thermal inertias on Europa range between 40 and \replaced{200}{300} $J/(m^2 \cdot K \cdot s^{1/2})$, with the lowest average values on the sub-Jovian hemisphere and the highest average values between the leading and anti-Jovian hemispheres. However, it is important to note that some patterns interpreted to be thermal inertia in this map may, in reality, be due to patterns in emissivity or albedo. In fact, this method does not completely eliminate residuals in all regions of overlap between the images, and minor discrepancies (primarily $\leq$ 3 K) remain, \edit1{implying some underlying differences in emissivity or albedo from those values used}. 

We extrapolate our albedos from relatively few \textit{Voyager} measurements, and this extrapolation may not work equally well for all locations on the surface. As mentioned above, the scatter between our best-fit albedo model and the measured albedo tie-points is +/-0.03, which results in adjustments to the best-fit local thermal inertia between 10--\edit1{60}\% for most locations on the surface. However, we do not see any obvious correlation between residuals, derived thermal inertias, and albedo. Furthermore, this map does produce a locally elevated thermal inertia surrounding the crater Pwyll, which is consistent with nighttime PPR data of the same region \citep{Spencer1999, Trumbo2017alma} and the tendency of crater ejecta to exhibit elevated thermal inertia elsewhere in the Solar System \citep[e.g.][]{Mellon2000,Hayne2017}. 

\edit1{However, we can take a complementary approach and assume that the residuals of Figure \ref{fig:residuals} are instead due to variations in the surface emissivity at our ALMA wavelength of 1.3 mm. Fixing the thermal inertia at our global best-fit value of 95 $J/(m^2 \cdot K \cdot s^{1/2})$ and letting the emissivity vary at each point produces a similar map of emissivity at 1.3 mm (Figure \ref{fig:map}) that alternatively fits the data with residuals of comparable magnitude in areas of overlap. Thus, the residuals of Figure \ref{fig:residuals} likely represent a combination of thermal inertia and emissivity variations.} \edit2{Indeed, we find both the fitted thermal inertia and emissivity ranges to be physically plausible. All derived thermal inertias are less than that of solid water ice, and the low end of our fitted thermal inertia range is consistent with thermal inertias derived for the Saturnian satellites \citep{Howett2010}. Similarly, the depressed millimeter emissivities we derive are consistent with those derived for Kuiper Belt and trans-Neptunian objects \citep{BrownButler2017, LellouchEtAl2017}.}

The most striking feature of \edit1{both maps} is the localized region of either highly elevated thermal inertia or low emissivity (deep blue in Figure \ref{fig:map}) near 90$\degree$ W and 23$\degree$ N, which coincides with the coldest spot in both the sub-Jovian to leading hemisphere and leading to anti-Jovian hemisphere images in Figure \ref{fig:residuals}. The fact that we see this anomaly twice in two different images and at two different times of day strongly suggests that it corresponds to a region that is truly distinct in its thermal properties. \edit1{While the maps suggest that the anomaly could either be explained by an elevated thermal inertia or a low emissivity, the morning temperature measurement at this location is too cold to be fit satisfactorily by thermal inertia alone, and both measurements are best fit by a moderate thermal inertia of 80 $J/(m^2 \cdot K \cdot s^{1/2})$ and a low emissivity of 0.66. A locally high thermal inertia may indicate a region of larger average particle size or of greater transparency to sunlight. A locally depressed emissivity may also be related to particle size (and therefore subsurface scattering properties) or may indicate a compositional difference resulting in a distinct spectral emissivity or more transparent surface at the ALMA wavelength \citep[e.g.][]{LellouchEtAl2016, LellouchEtAl2017}.} 

This region coincides with the location of highest water ice abundance mapped by \citet{BrownHand2013} and the location suggested to have the most crystalline water ice by \citet{Ligier2016}. However, it does not correspond to any particular geologic unit \citep{Leonard2017} or any unusual visible morphological or albedo features \citep{USGSmap}. The anomaly is located within a region of relatively low-resolution imaging and was not covered by the \edit1{published} \textit{Galileo} PPR \edit1{data}, so it is possible that a corresponding morphological, geologic, or thermal feature was simply not seen by \textit{Galileo}. For instance, one might imagine that recent resurfacing, perhaps via melt-through or diapirism, both of which are proposed explanations for nearby Murias Chaos at 84$\degree$W and 22$\degree$N \citep{Figueredo2002, Fagents2003}, could have resulted in a region of \edit1{recently extruded material with corresponding morphological expressions and different thermal properties than the surrounding regolith.} It is also interesting to note that the anomaly is almost directly antipodal to Pwyll, the largest fresh crater on Europa at 271$\degree$W and  25$\degree$S. Antipodal focusing of impact ejecta has been suggested as a potential explanation for a high thermal inertia deposit and corresponding geologic features on the Moon that are approximately antipodal to Tycho crater \citep{Bandfield2017, Hayne2017}. However, Tycho was likely the result of a more powerful impact, and is roughly three times larger than Pwyll in diameter. Unfortunately, without new, high-resolution imaging, the potential association of this feature with unique geology will likely remain an open question.

Curiously, with the exception of Pwyll crater, our analysis has not identified any clear correlations between the potential thermal inertia values and geologic or compositional units. Although, this is not necessarily surprising, if one draws analogy to the Moon, where thermal inertia reliably follows impacts, but not major geologic units \citep{Hayne2017}. Furthermore, unlike Mimas and Tethys \citep{Howett2011Mimas,Howett2012Tethys}, and despite its location within Jupiter's magnetosphere, Europa also appears to carry no obvious thermal inertia signature of high-energy electron bombardment. On Mimas and Tethys, the thermal inertia anomalies are associated with the leading hemisphere bombardment of electrons with energies $\geq$ 1 MeV (the energy needed for movement against the co-rotation direction of Saturn's magnetosphere) \citep{Howett2011Mimas, Howett2012Tethys}. It is possible that Europa lacks such a signature because the energy threshold for movement against co-rotation is much higher in Jupiter's magnetosphere ($\sim$25 MeV) \citep{Paranicas2007, Paranicas2009}, so electrons with energies $>$ 1 MeV bombard both the leading and trailing hemispheres of Europa.

\edit1{Ideally, we would include \textit{Galileo} PPR data in this analysis. However, while our modeling approach reproduces the ALMA images quite well, it does not produce the same quality fits to the \textit{Galileo} daytime data. When we try to incorporate the \textit{Galileo} daytime data into our analysis, we find that the \textit{Galileo} and ALMA data seem to prefer drastically different thermal properties, even in areas of overlap, such that the inclusion of the \textit{Galileo} PPR data leaves residuals that appear systematic. In fact, when we attempt to model the \textit{Galileo} PPR data alone, as we have done with the ALMA dataset, we still obtain systematic residuals, which are only eliminated if we allow albedo and thermal inertia to vary simultaneously. This approach, however, requires bolometric albedo patterns over much of the surface that show little correspondence with the \textit{Voyager}/\textit{Galileo} mosaic or our ALMA images. We suspect this may be the result of unexplained systematics in the PPR data, rather than real properties of Europa's surface. As we are unable to explain why the daytime PPR data is inconsistent with the ALMA daytime data in areas of overlap, we opt to forego using any PPR data and instead focus on our self-consistent ALMA dataset.} Despite this, our analysis does reproduce the high thermal inertia near Pwyll crater that was derived using both ALMA data and \textit{Galileo} PPR nighttime data \citep{Trumbo2017alma, RathbunEtAl2017}. If we apply the same minor albedo adjustment near Pwyll as \citet{Trumbo2017alma}, which is within our albedo errors, we arrive at \edit1{a similar} thermal inertia using only the ALMA data. 

\section{Conclusions}\label{sec:conclusions}
We obtained four ALMA thermal observations of Europa, which together cover the entire surface and reveal significant thermal structure. Using a globally homogenous, one-dimensional thermal model and a bolometric albedo map constructed from \textit{Voyager} measurements, we are able to reproduce much of this structure well, indicating that it is primarily a product of bolometric albedo variation across the surface and the passive absorption and re-emission of sunlight. However, despite the similarity of the data and model images, there are localized disagreements, which may indicate variability in the surface thermophysical properties. We examine the possibility that these discrepancies can be explained by local thermal inertia variations and construct a corresponding thermal inertia map, assuming a globally homogenous \edit1{millimeter} emissivity. The map suggests typical values of the surface thermal inertia ranging from 40 to \edit1{300} $J/(m^2 \cdot K \cdot s^{1/2})$, with the lowest thermal inertias on the sub-Jovian hemisphere and the highest between the leading and anti-Jovian hemispheres. \edit1{We also construct a complementary map of emissivity at our ALMA wavelength (1.3 mm), assuming a globally homogenous thermal inertia, which suggests emissivities of 0.67 -- 0.84.} We find little correlation \edit1{of thermal properites} with geology or composition and few noteworthy anomalies, with the exception of an elevated thermal inertia surrounding Pwyll crater and a region of \edit1{low emissivity} or extremely elevated thermal inertia near 90$\degree$W and 23$\degree$N on the leading hemisphere, in a region of relatively low-quality \textit{Galileo} imaging. This leading hemisphere location corresponds to the region spectroscopically determined to be the iciest (and potentially most crystalline) on the surface. However, it does not coincide with any unique geologic or morphological features, nor was it it covered by the \textit{Galileo} PPR. Thus, while we suggest that the area \edit1{is distinct in its thermal properties}, we can only speculate as to its origins. Future ALMA observations will provide measurements of each location on the surface at other times of day, which will allow for better constraint on the surface thermal properties and, thus, their potential influences. 

\acknowledgements
This paper makes use of the following ALMA data: ADS/JAO.ALMA\#2015.1.01302.S. ALMA is a partnership of ESO (representing its member states), NSF (USA) and NINS (Japan), together with NRC (Canada), MOST and ASIAA (Taiwan), and KASI (Republic of Korea), in cooperation with the Republic of Chile. The Joint ALMA Observatory is operated by ESO, AUI/NRAO and NAOJ. The National Radio Astronomy Observatory is a facility of the National Science Foundation operated under cooperative agreement by Associated Universities, Inc. This research was supported by Grant 1313461 from the National Science Foundation, as well as by a NASA Earth and Space Science Fellowship. The authors thank John R. Spencer, Julie A. Rathbun, James T. Keane, and Katherine R. de Kleer for useful conversations. \edit1{The authors would also like to thank the anonymous referee, whose insightful comments led to the improvement of this manuscript.}

\software{ALMA pipeline \citep{2014_Muders}, CASA \citep{McMullinEtAl2007}, AIPS \citep{aips}}


\begin{thebibliography}{}
\expandafter\ifx\csname natexlab\endcsname\relax\def\natexlab#1{#1}\fi

\bibitem[{Bandfield {et~al.}(2017)Bandfield, Cahill, Carter, Neish, Patterson,
  Williams, \& Paige}]{Bandfield2017}
Bandfield, J.~L., Cahill, J. T.~S., Carter, L.~M., {et~al.} 2017, Icarus, 283,
  282

\bibitem[{Barr \& Showman(2009)}]{BarrShowman2009}
Barr, A.~C., \& Showman, A.~P. 2009, Europa, ed. R.~T. Pappalardo, W.~B.
  McKinnon, \& K.~Khurana (Tucson, AZ: The University of Arizona Press),
  405--430

\bibitem[{Bridle \& Greisen(1994)}]{aips}
Bridle, A.~H., \& Greisen, E.~W. 1994

\bibitem[{Brown \& Butler(2017)}]{BrownButler2017}
Brown, M.~E., \& Butler, B.~J. 2017, The Astronomical Journal, 154, 19

\bibitem[{Brown \& Hand(2013)}]{BrownHand2013}
Brown, M.~E., \& Hand, K.~P. 2013, The Astronomical Journal, 145, 110

\bibitem[{Brown \& Matson(1987)}]{Brown1987}
Brown, R.~H., \& Matson, D.~L. 1987, Icarus, 72, 84

\bibitem[{{Butler} \& {Bastian}(1999)}]{1999_Butler}
{Butler}, B., \& {Bastian}, T.~S. 1999, in Astronomical Society of the Pacific
  Conference Series, Vol. 180, Synthesis Imaging in Radio Astronomy II, ed.
  G.~B. {Taylor}, C.~L. {Carilli}, \& R.~A. {Perley}, 625--656

\bibitem[{Butler(2012)}]{2012_Butler}
Butler, B.~J. 2012, {ALMA Memo 594}

\bibitem[{Carlson {et~al.}(2002)Carlson, Anderson, Johnson, Schulman, \&
  Yavrouian}]{Carlson2002}
Carlson, R.~W., Anderson, M.~S., Johnson, R.~E., Schulman, M.~B., \& Yavrouian,
  A.~H. 2002, Icarus, 157, 456

\bibitem[{Carlson {et~al.}(1999)Carlson, Johnson, \& Anderson}]{Carlson1999}
Carlson, R.~W., Johnson, R.~E., \& Anderson, M.~S. 1999, Science, 286, 97

\bibitem[{Collins \& Nimmo(2009)}]{Collins2009}
Collins, G., \& Nimmo, F. 2009, Europa, ed. R.~T. Pappalardo, W.~B. McKinnon,
  \& K.~Khurana (Tucson, AZ: The University of Arizona Press), 259--282

\bibitem[{{Cornwell} \& {Fomalont}(1999)}]{1999_Cornwell}
{Cornwell}, T., \& {Fomalont}, E.~B. 1999, in Astronomical Society of the
  Pacific Conference Series, Vol. 180, Synthesis Imaging in Radio Astronomy II,
  ed. G.~B. {Taylor}, C.~L. {Carilli}, \& R.~A. {Perley}, 187--199

\bibitem[{Fagents(2003)}]{Fagents2003}
Fagents, S.~A. 2003, Journal of Geophysical Research: Planets, 108

\bibitem[{Figueredo {et~al.}(2002)Figueredo, Chuang, Rathbun, Kirk, \&
  Greeley}]{Figueredo2002}
Figueredo, P.~H., Chuang, F.~C., Rathbun, J.~A., Kirk, R.~L., \& Greeley, R.
  2002, Journal of Geophysical Research: Planets, 107

\bibitem[{Fischer {et~al.}(2015)Fischer, Brown, \& Hand}]{Fischer2015}
Fischer, P.~D., Brown, M.~E., \& Hand, K.~P. 2015, The Astronomical Journal,
  150, 164

\bibitem[{Grundy {et~al.}(2007)Grundy, Buratti, Cheng, Emery, Lunsford,
  McKinnon, Moore, Newman, Olkin, Reuter, {et~al.}}]{Grundy2007}
Grundy, W.~M., Buratti, B.~J., Cheng, A.~F., {et~al.} 2007, Science, 318, 234

\bibitem[{Hayne \& Aharonson(2015)}]{Hayne2015}
Hayne, P.~O., \& Aharonson, O. 2015, Journal of Geophysical Research: Planets,
  120, 1567

\bibitem[{Hayne {et~al.}(2017)Hayne, Bandfield, Siegler, Vasavada, Ghent,
  Williams, Greenhagen, Aharonson, Elder, Lucey, {et~al.}}]{Hayne2017}
Hayne, P.~O., Bandfield, J.~L., Siegler, M.~A., {et~al.} 2017, Journal of
  Geophysical Research: Planets, 122, 2371

\bibitem[{Howett {et~al.}(2012)Howett, Spencer, Hurford, Verbiscer, \&
  Segura}]{Howett2012Tethys}
Howett, C. J.~A., Spencer, J.~R., Hurford, T., Verbiscer, A., \& Segura, M.
  2012, Icarus, 221, 1084

\bibitem[{Howett {et~al.}(2010)Howett, Spencer, Pearl, \& Segura}]{Howett2010}
Howett, C. J.~A., Spencer, J.~R., Pearl, J., \& Segura, M. 2010, Icarus, 206,
  573

\bibitem[{Howett {et~al.}(2011)Howett, Spencer, Schenk, Johnson, Paranicas,
  Hurford, Verbiscer, \& Segura}]{Howett2011Mimas}
Howett, C. J.~A., Spencer, J.~R., Schenk, P., {et~al.} 2011, Icarus, 216, 221

\bibitem[{Kattenhorn \& Hurford(2009)}]{Kattenhorn2009}
Kattenhorn, S.~A., \& Hurford, T. 2009, Europa, ed. R.~T. Pappalardo, W.~B.
  McKinnon, \& K.~Khurana (Tucson, AZ: The University of Arizona Press),
  199--236

\bibitem[{Lellouch {et~al.}(2016)Lellouch, Santos-Sanz, Fornasier, Lim,
  Stansberry, Vilenius, Kiss, M{\"u}ller, Marton, Protopapa,
  {et~al.}}]{LellouchEtAl2016}
Lellouch, E., Santos-Sanz, P., Fornasier, S., {et~al.} 2016, Astronomy and
  Astrophysics, 588, A2

\bibitem[{Lellouch {et~al.}(2017)Lellouch, Moreno, M{\"u}ller, Fornasier,
  Santos-Sanz, Moullet, Gurwell, Stansberry, Leiva, Sicardy,
  {et~al.}}]{LellouchEtAl2017}
Lellouch, E., Moreno, R., M{\"u}ller, T., {et~al.} 2017, Astronomy and
  Astrophysics, 608, A45

\bibitem[{Leonard {et~al.}(2017)Leonard, Patthoff, Senske, Collins, Bunte, \&
  Doggett}]{Leonard2017}
Leonard, E.~J., Patthoff, D.~A., Senske, D.~A., {et~al.} 2017, in LPSC,
  Vol.~48, 2357

\bibitem[{Ligier {et~al.}(2016)Ligier, Poulet, Carter, Brunetto, \&
  Gourgeot}]{Ligier2016}
Ligier, N., Poulet, F., Carter, J., Brunetto, R., \& Gourgeot, F. 2016, The
  Astronomical Journal, 151, 163

\bibitem[{McCord {et~al.}(1998)McCord, Hansen, Clark, Martin, Hibbitts, Fanale,
  Granahan, Segura, Matson, Johnson, Carlson, Smythe, \&
  Danielson}]{McCord1998}
McCord, T.~B., Hansen, G.~B., Clark, R.~N., {et~al.} 1998, Journal of
  Geophysical Research: Planets, 103, 8603

\bibitem[{McEwen(1986)}]{McEwen1986}
McEwen, A.~S. 1986, Journal of Geophysical Research: Solid Earth, 91, 8077

\bibitem[{McMullin {et~al.}(2007)McMullin, Waters, Schiebel, Young, \&
  Golap}]{McMullinEtAl2007}
McMullin, J.~P., Waters, B., Schiebel, D., Young, W., \& Golap, K. 2007, in
  Astronomical data analysis software and systems XVI, Vol. 376, 127

\bibitem[{Mellon {et~al.}(2000)Mellon, Jakosky, Kieffer, \&
  Christensen}]{Mellon2000}
Mellon, M.~T., Jakosky, B.~M., Kieffer, H.~H., \& Christensen, P.~R. 2000,
  Icarus, 148, 437

\bibitem[{Mitri \& Showman(2005)}]{MitriShowman2005}
Mitri, G., \& Showman, A.~P. 2005, Icarus, 177, 447

\bibitem[{Muders {et~al.}(2014)Muders, Wyrowski, Lightfoot, Williams, Nakazato,
  Kosugi, Davis, \& Kern}]{2014_Muders}
Muders, D., Wyrowski, F., Lightfoot, J., {et~al.} 2014, in Astronomical Data
  Analysis Software and Systems XXIII, ed. N.~Manset \& P.~Forshay, 383--386

\bibitem[{Paranicas {et~al.}(2009)Paranicas, Cooper, Garrett, Johnson, \&
  Sturner}]{Paranicas2009}
Paranicas, C., Cooper, J.~F., Garrett, H.~B., Johnson, R.~E., \& Sturner, S.~J.
  2009, Europa, ed. R.~T. Pappalardo, W.~B. McKinnon, \& K.~Khurana (Tucson,
  AZ: The University of Arizona Press), 529--544

\bibitem[{Paranicas {et~al.}(2007)Paranicas, Mauk, Khurana, Jun, Garrett,
  Krupp, \& Roussos}]{Paranicas2007}
Paranicas, C., Mauk, B.~H., Khurana, K., {et~al.} 2007, Geophysical Research
  Letters, 34, doi:10.1029/2007gl030834

\bibitem[{Pearl \& Sinton(1982)}]{Pearl1982}
Pearl, J.~C., \& Sinton, W.~M. 1982, in Satellites of Jupiter, ed. D.~Morrison,
  724--755

\bibitem[{Prockter \& Patterson(2009)}]{Prockter2009}
Prockter, L.~M., \& Patterson, G.~W. 2009, Europa, ed. R.~T. Pappalardo, W.~B.
  McKinnon, \& K.~Khurana (Tucson, AZ: The University of Arizona Press)

\bibitem[{Putzig {et~al.}(2005)Putzig, Mellon, Kretke, \&
  Arvidson}]{Putzig2005}
Putzig, N.~E., Mellon, M.~T., Kretke, K.~A., \& Arvidson, R.~E. 2005, Icarus,
  173, 325

\bibitem[{Rathbun {et~al.}(2010)Rathbun, Rodriguez, \& Spencer}]{Rathbun2010}
Rathbun, J.~A., Rodriguez, N.~J., \& Spencer, J.~R. 2010, Icarus, 210, 763

\bibitem[{Rathbun \& Spencer(2017)}]{RathbunEtAl2017}
Rathbun, J.~A., \& Spencer, J.~R. 2017, in AGU Fall Meeting Abstracts

\bibitem[{Spencer(1987)}]{Spencer1987}
Spencer, J.~R. 1987, PhD thesis, University of Arizona, Tuscon, AZ

\bibitem[{Spencer(1990)}]{Spencer1990}
---. 1990, Icarus, 83, 27

\bibitem[{Spencer {et~al.}(1989)Spencer, Lebofsky, \& Sykes}]{Spencer1989}
Spencer, J.~R., Lebofsky, L.~A., \& Sykes, M.~V. 1989, Icarus, 78, 337

\bibitem[{Spencer {et~al.}(1990)Spencer, Shure, Ressler, Goguen, Sinton,
  Toomey, Denault, \& Westfall}]{Spencer1990Io}
Spencer, J.~R., Shure, M.~A., Ressler, M.~E., {et~al.} 1990, Nature, 348, 618

\bibitem[{Spencer {et~al.}(1999)Spencer, Tamppari, Martin, \&
  Travis}]{Spencer1999}
Spencer, J.~R., Tamppari, L.~K., Martin, T.~Z., \& Travis, L.~D. 1999, Science,
  284, 1514

\bibitem[{Spencer {et~al.}(2006)Spencer, Pearl, Segura, Flasar, Mamoutkine,
  Romani, Buratti, Hendrix, Spilker, \& Lopes}]{Spencer2006}
Spencer, J.~R., Pearl, J.~C., Segura, M., {et~al.} 2006, Science, 311, 1401

\bibitem[{Thompson {et~al.}(2001)Thompson, Moran, \& Swenson}]{2001_Thompson}
Thompson, A.~R., Moran, J.~M., \& Swenson, G.~W. 2001, Interferometry and
  Synthesis in Radio Astronomy, 2nd Edition (New York, New York:
  Wiley-Interscience)

\bibitem[{Trumbo {et~al.}(2017)Trumbo, Brown, \& Butler}]{Trumbo2017alma}
Trumbo, S.~K., Brown, M.~E., \& Butler, B.~J. 2017, The Astronomical Journal,
  154, 148

\bibitem[{Urquhart \& Jakosky(1996)}]{Urquhart1996}
Urquhart, M.~L., \& Jakosky, B.~M. 1996, Journal of Geophysical Research:
  Planets, 101, 21169

\bibitem[{USGS(2002)}]{USGSmap}
USGS. 2002, Controlled photomosaic map of Europa, Je 15M CMN: U.S. Geological
  Survey Investigations Series I-2757, ,

\end{thebibliography}
\end{document}